\documentclass[12pt,article,floats,superscriptaddress]{revtex4}

\usepackage{graphicx}

\newcommand{\be}{\begin{equation}}
\newcommand{\ee}{\end{equation}}
\newcommand{\bea}{\begin{eqnarray}}
\newcommand{\eea}{\end{eqnarray}}

\begin{document}

\preprint{AIP/123-QED}

\title{Characteristic Sign Change of the Magnetoresistance of Strongly Correlated GaAs Two-dimensional Holes}
\thanks{email:jianhuang@wayne.edu}

\author{Jian Huang}
\affiliation{%
Department of Physics and Astronomy, Wayne State University, Detroit, MI 48201, USA\\}%
\author{L. N. Pfeiffer}%
\author{K. W. West}%
\affiliation{%
Department of Electrical Engineering, Princeton University, Princeton, NJ 08544}%

\date{\today}

\begin{abstract}

High quality strongly correlated two-dimensional (2D) electron systems at low temperatures $T\rightarrow 0$ exhibits an apparent metal-to-insulator transition (MIT) at a large $r_s$ value around 40. We have measured the magnetoresistance of 2D holes in weak perpendicular magnetic field in the vicinity of the transition for a series of carrier densities ranging from $0.2-1.5\times10^{10}$ $cm^{-2}$. The sign of the magnetoresistance is found to be charge density dependent: in the direction of decreasing density, the sign changes from being positive to negative across a characteristic value that coincides with the critical density of MIT.
\end{abstract}

\pacs{Valid PACS appear here}
\keywords{GaAs two-dimensional hole(2DH)}
\maketitle


The transport of a two-dimensional electron system (2D) is strongly influenced by electron-electron interaction which is expected to fundamentally alter the manybody ground states from the usual Fermi Liquid (FL)~\cite{Anderson}. Extensive investigations targeting understanding the interaction-driven effects have been made in both zero and high magnetic fields. In the regime of the Fraction Quantum Hall Effect (FQHE)~\cite{Tsui} observed in the high magnetic (B) fields, there exists peculiar insulating states~\cite{FWC,reentrant-hi}, i.e. those appearing near filling factors 1/3 and 1/5, that are suspected as pinned Wigner Crystals~\cite{wc} driven by prominent interaction. On the other hand, in the absence of the B-field, the apparent 2D metal-to-insulator transition (MIT)~\cite{mit} has been observed with the presence of non-perturbative or even dominating interaction. Specifically for the large $r_s$ regime ($\geq$37~\cite{wc1}), MIT has been considered as a possible manifestation of the liquid to Wigner Crystal (WC) transition~\cite{Yoon}. The question on whether the zero field MIT and the high field reentrant phenomena share the same origin has been studied both experimentally and theoretically though no definitive conclusion has been reached. With improved sample quality, higher purity and more dilute 2D systems are now available~\cite{jian-1}. The apparent MIT has been observed with a transition around $r_s\sim37$. Even for 2D holes with larger effective mass, this means a density range down to at least $4-5\times10^{9}$ cm$^{-2}$ for which the high field FQHE measurement becomes difficult. However, the low-field measurement is still viable. This opens up questions on whether there are certain low-field characteristics that captures the influences due to interaction linking to the MIT. On another note, since a large B-field significantly alters the individual electron wavefunctions that are key to the formation of the many-body states, electron wavefunctions are preserved in zero and low B-fields.

The quantum corrections to the magnetoresistance $\rho_{xx}(B)$ due to weak localization (WL) and weak antilocalization (WAL) are well studied for metals. In systems with moderate interaction, such as dilute GaAs systems, WL and WAL are suppressed and the sign of the $\rho_{xx}(B)$ depends on the interplay between backscattering due to disorder, which results in WL with negative $\rho_{xx}(B)$, and the spin-orbit interaction (SOI), which gives rise to positive $\rho_{xx}(B)$~\cite{AKS-review,zna,mitGaAs,mills}. Especially, the study of SOI in GaAs 2D holes~\cite{interbandE,SOI2,SOI1} has attracted much interests for both fundamental and practical reasons (such as spintronics applications). However, in systems with more dilute charge concentrations, e.g. $10^{9}$ cm$^{-2}$ range, how $\rho_{xx}(B)$ is influenced by interaction, especially in the Wigner crystal regime is not well understood.

We report a study of the magnetoresistance $\rho_{xx}$ of 2D holes in HIGFETs (heteroheterojunction-insulated-gate field-effect-transistor)~\cite{kane} in a weak perpendicular B-field for several fixed dilute charge densities across the vicinity of the MIT from $2\times10^{9}$ cm$^{-2}$ to $1.5\times10^{10}$ cm$^{-2}$. Without any intentional doping, disorders in HIGFETs are effectively suppressed. Such systems have been previously demonstrated with ultra-high quality and altra-low carrier concentrations (down to $6\times10^{8}$ cm$^{-2}$)~\cite{jian-1}. Remarkably, the transport, even for $r_s$ values above 40, is qualitatively nonactivated~\cite{jian-image}, supporting a more genuine interaction-driven many-body state in the WC regime. The findings are summarized below.

For the metallic side at higher densities, Shubnikov–de Haas (SdH) oscillations are observed between 0.1 to 0.3 Tesla and the Fourier analysis reveals a single peak. Meanwhile, there also arises interaction-driven effects that are absent from the high density systems measured in weak $B_{\perp}$: a substantial rise of $\rho_{xx}$ observed before filling factor $\nu=1$ echoes with a recent observation in relation to a possible insulating behavior due to interaction-driven reentrant effect~\cite{reentrant}. Moverover, the magnetoresistance $\rho_{xx}$ is positive for this higher density range which is consistent with previous findings. However, the sign of the magnetoresistance is observed to be hole-density (or interaction) dependent. For the first time, we have observed a sign change of the magnetoresistance $\rho_{xx}$ from being positive to negative as the charge density is decreased across some characteristic charge density $p^*$. Moreover $p^*$ coincides reasonably well with the critical density of zero-field MIT. In another words, the sign of $\rho_{xx}$ becomes negative when the system crosses into the insulating regime of MIT. Therefore, the zero field MIT corresponds to the sign change of the magnetoresistance in weak field. In addition, the sign change of $\rho_{xx}$ is relatively abrupt around the critical density, suggesting a transition around $r_s\sim37$.

The GaAs (100) HIGFETs contain no doping and the charges are solely capacitively induced at the GaAa/AlGaAs heterjunction through a metal gate located 600 nm away from the 2D channel. The sample preparation details are provided in Ref.~\cite{jian-4}. The measurement was made in a dilution refrigerator with the sample placed in the 3He/4He mixing chamber. The density of the 2D holes was determined by measuring the quantum oscillations of the magnetoresistance and the Hall resistance in a perpendicular B-field. The result from the $T$-dependence of the resistivity in zero magnetic field indicates a critical density of MIT is $p_c\sim4.2\times10^{9}$ $cm^{-2}$ which corresponds to $r_s\sim 38$ if taking $m^*=0.3m_0$. The following are the magnetoresistance results for various fixed charge densities measured in weak perpendicular magnetic field swept at 100 Gauss/min.

\begin{figure}[b]
\vspace{-35pt}
\includegraphics[totalheight=3.0in,trim=0.4in 0.10in 0.20in 0in]{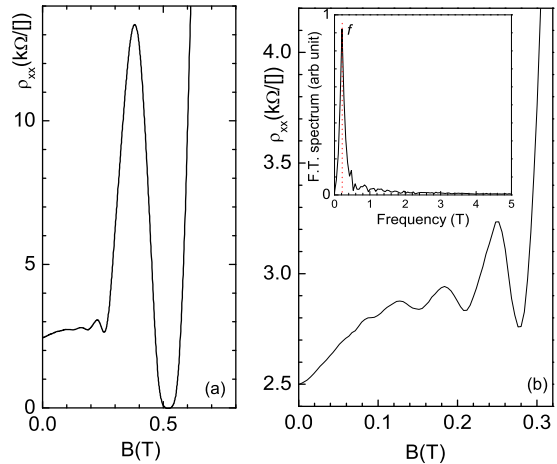}
\vspace{-20pt}
\caption{\label{fig:Rxx} (a) $\rho_{xx}$ vs. B for $p=1.2\times10^{10}$ cm$^{-2}$ at $T$= 85 mK. (b) SdH oscillations seen in a zoom-in view of (a) at low fields. Inset: Fourier spectrum of the shown $\rho_{xx}(B)$.}
\vspace{-10pt}
\end{figure}

Fig.\ref{fig:Rxx}(a) shows the behavior of the quantum oscillations of the magnetoresistance in a perpendicular $B$ field up to 0.6 Tesla measured at $T=85$ mK. The carrier density determined by $\rho_{xx}(B)$ and the Hall resistance ($\rho_{xy}$) is $1.24\times10^{10}$ cm$^{-2}$. The carrier mobility is $\sim300,000$ cm$^{2}/(V\cdot s)$ with $\rho_{xx}=2.35$ k$\Omega/\sim0.09h/e^2$ at $B=0$. The corresponding $r_s$ value is approximately 33 by taking the effective mass of 0.45 $m_0$~\cite{mass}. The SdH oscillations are better seen in the blowup figure [Fig.~\ref{fig:Rxx}(b)] where the amplitude of the the oscillations becomes visible around 0.03 T and rises with increasing $B$ only up to 0.3 T before $\rho_{xx}$ exhibits a peak right before filling factor $\nu=1$ appearing at 0.53 T.

Note first that SdH only shows up in a very narrow range due to low carrier density which is 5-10 \% of those measured previously~\cite{SOI1,SOI2}. The Fourier analysis plotted in the inset of Fig.~\ref{fig:Rxx}(b) shows only one characteristic frequency peak around $f=0.25$T with a width of $\sim$0.12 T. The width of the $f$ peak indicates a resolution of resolving density differences between two subbands to be $2.5\times10^{9}$ $cm^{-2}$. The corresponding charge density to the $f$ peak is approximately $6\times10^{9}$ $cm^{-2}$ which gives a total density of $1.2\times10^{10}$ $cm^{-2}$ with an error bar of $3\times10^{9}$ $cm^{-2}$, consistent with the density determined through Hall measurement. As experimentally demonstrated previously for high charge densities (in the $10^{11}$ $cm^{-2}$ range), SOI lifts the degeneracy of the heavy hole band and splits it into two spin subbands which are identified through SdH oscillations characterized by two different frequencies~\cite{SOI1,SOI2}. The single peak in FT analysis for our case indicates that the spin subbands are degenerate in this dilute charge limit, consistent with the estimate from Ref.~\cite{interbandE} stating that degeneracy occurs for $p<1\times10^{11}$ $cm^{-2}$. It worth noting that, for the very dilute $p$ limit, HH (heavy hole) and LH (light hole) bands mixing~\cite{mass} will eventually arrive. Here, the advantage for having a single (degenerate) band is that the interband scattering~\cite{interbandE} is insignificant and SOI plays the prime role of affecting the low field magnetoresistance.

\begin{figure}[t]
\vspace{-30pt}
\includegraphics[totalheight=4.8in,trim=0.4in 0.10in 0.10in 0in]{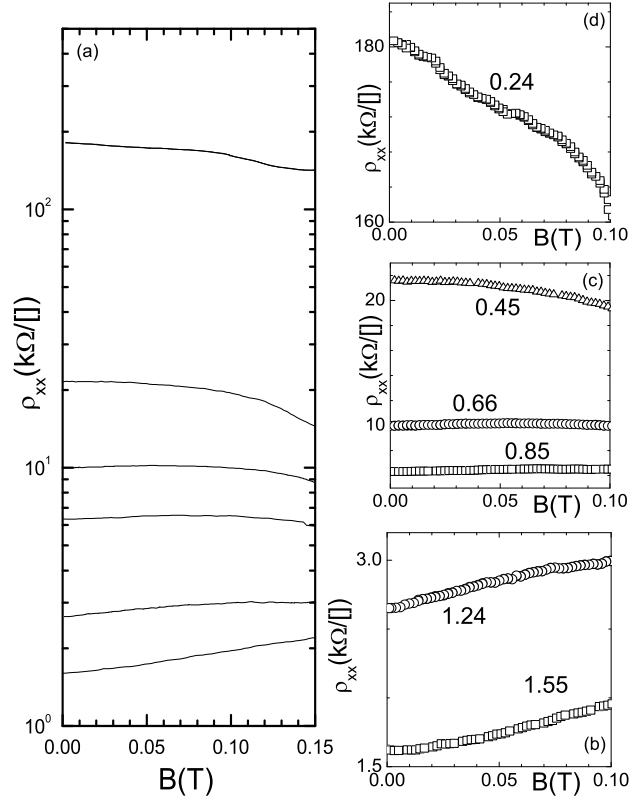}
\vspace{-20pt}
\caption{\label{fig:weakB} (a) $\rho_{xx}(B)$ in weak perpendicular B-field for various hole densities (from bottom up): 1.55, 1.24, 0.85, 0.66, 0.44, 0.25 $\times10^{10}$ cm$^{-2}$; (b), (c), and (d) are the blowup figures for various indicated densities in the unit of $10^{10}$ $cm^{-2}$.}
\vspace{-10pt}
\end{figure}

As also seen in Fig.~\ref{fig:Rxx}(a), $\rho_{xx}$ exhibits a peak around $B=0.37$T between filling factors 2 and the 14 k$\Omega$ peak resistivity, six times the value of $\rho_{xx}(B=0)$, which is overwhelmingly higher than the amplitudes of the SdH oscillations. This is in agreement with a recent observation made from doped $p$-GaAs quantum well (QW) systems in relation to the reentrant insulating phase (RIP) even possible Wigner Crystal phase~\cite{reentrant} around 0.2 T, as also related to the insulating phases, observed at lower fractional states for higher density systems~\cite{reentrant-hi}.

The positive magnetoresistance shown in figure~\ref{fig:Rxx}(b) has a slope of approximately $\sim$7.5 k$\Omega/$Tesla up to 0.05T, with an overall rise of $\rho_{xx}(B)$ is approximately 18\% of $\rho_{xx}(B=0)$. We note that in $p$-GaAs quantum well systems, negative $\rho_{xx}(B)$ was previously observed as a result of an interplay of the WL and WAL effects~\cite{mitGaAs,spivakgao}. The SOI demonstrated in undoped GaAs/AlGaAs heterostructures shows a stronger effect. The cause of the positive $\rho_{xx}(B)$ has been deemed as combined effects due to both inter-subbands scattering and SOI~\cite{SOI1}. Here, since the SdH result supports a degenerate band at this low density limit, we need to consider only the effect of SOI. The substantial increase of $\rho_{xx}(B)$ indicates a significant though not dominating SOI. An important question is how the effect of SOI evolves in the presence of progressively stronger interaction tuned via lowering $p$.

The magnetoresistance $\rho_{xx}(B,p)$ in weak $B_\bot$ field is shown in figure~\ref{fig:weakB}(a) for a series of fixed charge densities from $1.55\times10^{10}$ cm$^{-2}$ down to $0.24\times10^{10}$ cm$^{-2}$. The average charge spacing 2$a$, $a=1/\sqrt{\pi p}$, varies from 160nm to 300 nm. The corresponding $r_s$ value is from 25 to beyond 40 (assuming $m^*=0.2m_0$). The exact $r_s$ value for the lowest densities is hard to determine due to two factors. The first is the unknown effective mass~\cite{mass}. The second is the screening effect from the metallic gate~\cite{Jian-screening} which undercuts the long-ranged Coulomb effect. Usually, logarithmic growth of $\rho_{xx}(B)$ is expected without considering strong interaction. Since we are dealing with a very short field range and it also involves a change of sign of $\rho_{xx}(B)$, we used a linear fit to approximately obtain the slope up to 0.1 Tesla.

As shown in Fig.\ref{fig:weakB}, the slope of the magnetoresistance at lowest field is $p$ dependent: as $p$ is decreased, the positive slope of $\rho_{xx}(B)$ starts to decrease and becomes flat when $p$ reaches $\sim4\times10^{9}$ cm$^{-2}$. For lower $p$, however, $\rho_{xx}(B)$ becomes negative and $\rho_{xx}(B=0,p)$ also rises above $h/e^2$. We compare the results to a previous measurement with doped GaAs/AlGaAs heterostructures with a minimum charge density of $7\times10^{9}$ cm$^{-2}$ ($r_s<30$)~\cite{mills} which shows a severely suppressed negative $\rho_{xx}$ and the WL effect becomes negligible in the presence of strong interaction. Therefore, the change of signs of $\rho_{xx}(B)$ we observed around $p\sim4\times10^{9}$ cm$^{-2}$ points to a different origin, especially considering the negative $\rho_{xx}(B)$ for $r_s\geq37$.


\begin{figure}[b]
\vspace{-35pt}
\includegraphics[totalheight=3.2in,trim=0.3in 0.10in 0.20in 0in]{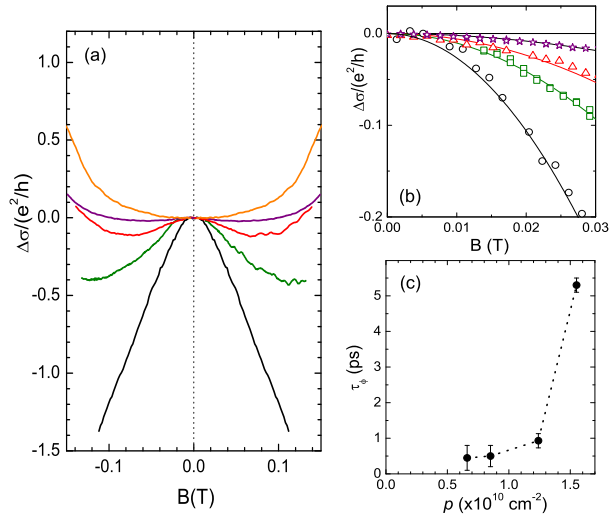}
\vspace{-20pt}
\caption{\label{fig:fit} (a) $\Delta\sigma(B)$ for various densities in low fields. (b) Fittings (solid lines) of $\Delta\sigma(B)$ based on the HLN theory. (c) Estimated $\tau_{\phi}$ according to the fittings in the limit of $\tau_{so}\gg\tau_{\phi}$}
\vspace{-10pt}
\end{figure}

Nevertheless, we fit the $\rho_{xx}(T)$ for the low field to the theoretical model~\cite{HLN}, developed by Hikami et al. for the metallic states of charges, to compare with previous results~\cite{mitGaAs,mills} for which the same fitting method were performed. Figure~\ref{fig:fit}(a) is the magnetoconductance $\Delta\sigma(B)$, $\sigma(B)$-$\sigma(B=0)$ in the units of $e^2/h$, which resembles the previous findings~\cite{SOI2} even though the tuning factor here is the carrier density at a fixed T. In the low field diffusive limit ($\tau_{so}\gg\tau\phi$), the correction to the magnetoconductance  $\Delta\sigma(B)=\sigma(B)-\sigma(B=0)$, depending mainly on $\tau_\phi$ (the inelastic scatter time), can be written as
\begin{eqnarray} \label{HLN}
\Delta\sigma(B)=-\frac{e^2}{\pi h}[\frac{1}{2}\Psi(\frac{1}{2}+\frac{B_\phi}{B})-\frac{1}{2}ln(\frac{B_\phi}{B})]
\end{eqnarray}
\noindent where $\Psi$ is the digamma function, $B_\phi$ is the effective field related to $\tau_\phi$ via $\tau_\phi=\frac{\hbar}{4eDB_\phi}$ with $D$ being the diffusion constant. The fittings are shown in figure~\ref{fig:fit}(b) for four different densities down to $6.6\times10^{9}$ cm$^{-2}$. The estimated $\tau_\phi$ as a function of the carrier density is shown in panel (c). For $p=1.55\times10^{10}$ cm$^{-2}$, $\tau_{\phi}\sim5.5$ $p$s, similar to the previous findings from 2D holes in GaAs quantum wells~\cite{mitGaAs}. Contrast to the T-dependence of $\tau_\phi$ demonstrated in Ref.~\cite{SOI2,mitGaAs}, the lowering of the density causes a more rapid drop from 5.5 ps to 1 ps for $p$ from $1.55$ to $1.24\times10^{10}$ cm$^{-2}$. This is consistent with the increasing {\it ee} interaction effect (as the $r_s$ value rises from 25 to 30) as well as the weakening of the screening effect~\cite{Tanaskovic,screening}. The fitting gets increasingly less satisfactory with decreasing $p$ as indicated with increasing error bars. And the slower decrease of $\tau_\phi$ with decreasing $p$ probably indicates the fitting is already no longer valid even for $p\sim1\times10^{10}$ cm$^{-2}$. For $p=4\times10^{9}$ cm$^{-2}$, the conductivity falls below $e^2/h$ and the fitting is no longer possible. The fitting further confirms the distinction between a strongly correlated system from a metal.

\begin{figure}[t]
\vspace{-20pt}
\includegraphics[totalheight=2.8in,trim=0.1in 0.10in 0.20in 0in]{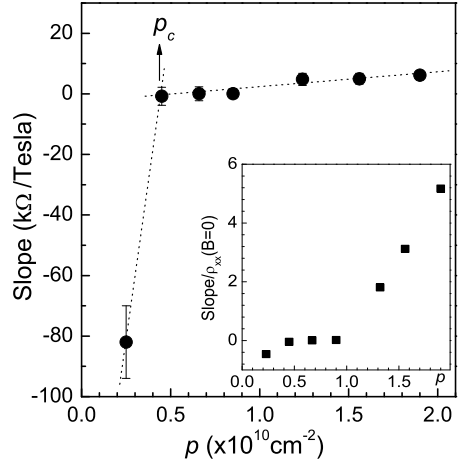}
\vspace{-10pt}
\caption{\label{fig:slope} Low field slope of $\rho_{xx}(B)$ showing a sign change as varying $p$ across the critical density $\sim4\times10^{9}$ $cm^{-2}$.}
\vspace{-15pt}
\end{figure}

For $r_s\geq37$, we have already demonstrated a nonactivated transport in the same high purity systems that suggests a strongly correlated liquid of possibly semiquantum nature, instead of an Anderson insulator~\cite{jian-image} or an (unmelted) Wigner solid. The 85 mK experimental temperature is close to the classical melting temperature $T_{melt}\sim E_{ee}/137\sim 120$ mK for $p=4\times10^{9}$ cm$^{-2}$. Quantum fluctuations further decrease the $T_{melt}$. In addition, the current excitation used for measurement further reduces the melting point through nonequilibrium effects~\cite{WC_helium,Jian_WC}. Thus, it is reasonable to consider it as a semiquantum liquid whose transport can be modeled as a hydrodynamic flow of carriers~\cite{Andreev}. Therefore, the "abrupt" sign change of $\sigma(B)$ at $p_c$ is likely a result of such a transition into a melted Wigner crystal. In order to observe a liquid to solid transition, both lower temperature and ultra-low excitations are required~\cite{Jian_WC}.

As for the characteristics of the $\rho_{xx}(B)$ sign change, first, it occurs at a characteristic density of $\sim4\times10^{9}$ cm$^{-2}$ that coincides with the critical density $p_c$ of MIT, and this critical density corresponds approximately to $r_s=38$ assuming $m^*=0.3m_0$~\cite{mass}. This correlates the zero-field MIT indicated by the sign change of $d\rho/dT$ ($T$ is the temperature) with the sign change of the $\rho_{xx}(B)$ in weak magnetic field into the Wigner Crystal regime. Second, the slope of $\rho_{xx}(B)$ with respect to the change of the charge density, as shown in figure~\ref{fig:slope}, is apparently abrupt around $p_c$. To approximately quantify the sign change of the magnetoresistance, we choose a linear fit within a short window of the field up to 0.1 T and the slope change is shown as a function of $p$. The slope for the negative $\rho_{xx}$ with respect to the change of the charge density is almost an order of magnitude greater than that for the positive $\rho_{xx}$ cases. The inset is the slope normalized by the $\rho_{xx}(B=0)$ values for each specific density which also confirms this abruptness around $p_c$. Both plots show a piece-wise behavior in correspondence to the MIT.


How a weak B field affects a Wigner crystal or melted Winger crystal is not well understood. In the FQHE regime, a B field is believed to enhance the crystallization of charges. In our case, the negative $\rho_{xx}(B)$ for $r_s>37$ in a weak field seems to indicate an opposite effect as if there is an increasing effect of melting or a decreasing pinning effect with increasing B. It is, intuitively, a possibility considering the influence of a B field on the dynamical depinning of a (melted) Winger crystal~\cite{nori}. Note that even for a melted Wigner crystal, disorder effect is still manifested through a viscous flow~\cite{spivakgao}. More theoretical work is needed for a better understanding. Meanwhile, striped Wigner Crystal phase~\cite{stripe} may also be relevant since in the lower density limit, the screening from the gate is expected to undercut both the long range Coulomb interaction and the bulk disorder potentials to $1/\sqrt{(2r)^2+(2d)^2}$ ($r$ is the spatial distance and $d$ is the distance from the channel to the metallic gate). This theory suggests further measurement ideas with L-shape hallbar which can distinguish transport behaviors along different directions.

To summarize, we have measured the magnetoresistance of ultradilute 2D holes down to $2\times10^{9}$ cm$^{-2}$ in weak perpendicular field and found a correspondence between a sign change of the magnetorsistance and the apparent MIT across the same critical density. Especially, the negative signs of the $\rho_{xx}(B)$ for $r_s\geq37$ suggests an unknown origin linked to strong interaction effects.

The support of the work is through NSF-1105183.

\bibliography{aipsamp}

\end{document}